# CPU-less parallel execution of lambda calculus in digital logic

Harry Fitchett[1] and Charles Fox[1]

[1]School of Engineering and Physical Science, University of Lincoln, UK.

**Abstract**

While transistor density is still increasing, clock speeds are not, motivating the search for new parallel architectures. One approach is to completely abandon the concept of CPU – and thus serial imperative programming – and instead to specify and execute tasks in parallel, compiling from programming languages to data flow digital logic. It is well-known that pure functional languages are inherently parallel, due to the Church-Rosser theorem, and CPU-based parallel compilers exist for many functional languages. However, these still rely on conventional CPUs and their von Neumann bottlenecks. An alternative is to compile functional languages directly into digital logic to maximize available parallelism. It is difficult to work with complete modern functional languages due to their many features, so we demonstrate a proof-of-concept system using lambda calculus as the source language and compiling to digital logic. We show how functional hardware can be tailored to a simplistic functional language, forming the ground for a new model of CPU-less functional computation. At the algorithmic level, we use a tree-based representation, with data localized within nodes and communicated data passed between them. This is implemented by physical digital logic blocks corresponding to nodes, and buses enabling message passing. Node types and behaviors correspond to lambda grammar forms, and beta-reductions are performed in parallel allowing branches independent from one another to perform transformations simultaneously. As evidence for this approach, we present an implementation, along with simulation results, showcasing successful execution of lambda expressions. This suggests that the approach could be scaled to larger functional languages. Successful execution of a test suite of lambda expressions suggests that the approach could be scaled to larger functional languages.

# 1 Introduction

Transistor density continues to increase, while clock speeds do not. As demand for computational power continues to increase, the pressure is thus now to find ways to make maximal use of parallelism [1]. While a few algorithms are inherently parallelizable at the algorithmic level – such as the neural networks run on GPUs for AI [2][3] – most are still expressed as traditional serial imperative programs, which make them difficult to accelerate.

There is still ongoing debate about the most effective ways to parallelize programs. Two general approaches are commonly discussed. The first is explicit parallelization, where programmers must efficiently design and optimize parallel programs [4] [5]. The second is implicit parallelization, where the system itself attempts to automatically determine the best way to parallelize the program [6]. Implicit parallelization is particularly challenging in imperative languages because it is difficult to track the necessary interactions between tasks. A common strategy is to decompose functions into smaller sub-functions, which can then be executed independently in parallel. Pure functional programs naturally fit this model since they avoid mutable state, making them much easier to parallelize implicitly [7][8].

Functional languages have existed since Church's 1936 lambda calculus [9]. Lambda calculus and derived languages benefit from the Church-Rosser theorem, which proves that if any terminating sequence of reductions exist, then all terminating sequences of executions – such as call-by-name, call-by-value, and eager execution strategies – will evaluate to the same expression. In particular, this means that all arguments to a function can be evaluated in parallel, along with the function itself using name placeholders until the argument evaluations are complete and can be substituted for them. Some functional programs are represented and processed as tree-like graphs by implementations, referred to as graph simplification or reduction, which can be parallelized as multiple nodes reducing together [10][11].

Functional languages based on lambda calculus such as Lisp, Clean, ML, and Haskell have implementations on current CPU-based hardware [12][13]. However, the von Neumann architecture, from which CPUs derive, was designed to process serial programs, and while modern CPUs include some parallel elements which can be exploited [7], individual CPU cores still require imperative machine code, so functional languages typically compile programs to such code, even for performing parallel graph reduction [14][15]. Much work has been invested in maximizing the efficacy of functional compilers.

Term graph rewriting uses refined logical steps in lambda calculus reduction in order to speed up a costly portion of graph reduction, rewriting [16]. Techniques such as Half Combustion – a type of lazy execution [4][17] – have been used to optimize compilers. PELCR, a lambda-based language compiler, uses these techniques to achieve a near 80% speedup compared to prior established methods [18][19][17]. Alternate, more abstract, expression types have been proposed to allow for predicting futures appearing in $\lambda^{as}$ these expressions can evaluate branches below them to skip unnecessary steps [19][17][20]. Other language implementations aim to reduce the number of nodes required to represent a tree, by adding expressions to represent abstract data types such as lists.

Following these decisions, hardware engineers have asked what is the most efficient way to represent and execute functional languages, under the particular constraints that hardware imposes [21]. Such hardware designs are split into two camps, based on localized or global data access. Global data access is the more conventional approach, in which data is stored in large globally accessible memory components. Any computational unit currently executing a function may access this shared memory to retrieve information needed for the execution, such as values or pointer information. This method has achieved relative success and wide-scale adoption because it works well with familiar imperative programming and von Neumann architectures, An example in current use is OpenCL. Unfortunately, accessing memory components costs time and when multiple computational units are all trying to access a shared memory component, read/write bottlenecks can become more noticeable. Without proper management, this can lead to unhandled race conditions, where data is read before it is ready, in these cases it is up to the programmer to organize memory access to avoid unpredictable results. In trivially parallel programs this risk is minimal. However, others may have more interaction complexity between their tasks, which can worsen these bottlenecks.

Alternatively, data is stored in local memory only accessible by a single or limited number processing unit. Localized solutions have been built by representing functions and task across a number of computational units. Examples include distributed, packet-based and transputer implementations [14][22][23][13], in which separate machines work simultaneously. These solutions bypass the read/write bottleneck by limiting the number of machines that can access a memory unit. Previously this has been accomplished through distributed solutions in which nodes send packets to one another to pass information.

All of these previous localized data architectures for parallel execution of functional languages rely on distributed computation based on von Neumann machines. In contrast, we explore the possibility of compiling a simple functional language – lambda calculus – directly into digital logic. Digital logic gates are inherently parallel and may remove the need for CPUs altogether. Connecting them with integrated circuit level wiring may also be faster than relying on distributed networking levels of computation.

## 2 Hardware Considerations

As with software, good hardware solutions need to be evaluated and optimized. In this study, we will assess solutions using three key metrics: time requirements, space efficiency, and energy costs. When representing common abstract data structures like arrays, lists, and trees in digital logic, they must be stored in larger memory components, often spread across multiple RAM addresses. To manage this, dedicated memory space is typically allocated, and special pointer registers are used to keep track of key values locations. Accessing, modifying, and storing data in these structures also takes time. Implementing a lambda reducer in digital logic adds further complexity. It requires intricate structures to represent lambda expressions and their relationships to one another. This section explores the challenges of representing lambda expressions and graph-like structures in digital logic and proposes possible solutions, evaluated using the time, space, and energy metrics introduced earlier.

## 2.1 Localization of Data

The two memory models (global and local) have their advantages and disadvantages. For the global memory model, where a single accessible memory component is used, any computational unit can access the pointer values of any expression. The main advantage here is space efficiency since there is no need for complex routing, data can be accessed simply by following a chain of pointers to the desired expression. However, this approach can become inefficient over time. Constantly querying the shared memory can lead to delays, and the centralized access creates a bottleneck. As more computational units try to read and write data, performance slows down and energy costs increases significantly. Therefore we use the second approach, localized memory which spreads data across multiple smaller, local memory components. Here, each unit accesses only a small, localized portion of memory. The main challenge with this setup is connectivity. Data stored outside a unit's local memory still needs to be accessed. This requires a complex network of buses and routing logic, which takes up more space. However, this approach can significantly reduce both time delays and heat costs. This project pushes the localized memory model to its limit by limiting a units memory and processing capabilities to store and resolve a singular lambda calculus expression. Individual computational units, which from now on will be referred to as 'nodes', will have access to a single expression's values, maximizing the performance benefits from the local memory model.

## 2.2 Node Connectivity

As we have discussed, working with localized data requires a structure that allows nodes to communicate relevant information efficiently. The traditional response would be to have the nodes share a bus. Then, based on a scheduling protocol nodes could receive and transmit when relevant. Unfortunately, during computation multiple nodes may wish to exchange data simultaneously and to account for this multiple shared buses will be required to preserve the parallel properties that make this solution desirable.

The challenge is finding the right balance between scalability (keeping things efficient and compact) and flexibility (allowing any node to connect where needed). To illustrate this, consider two extreme examples: First a highly scalable, low flexibility graph. Imagine a rigid graph where each node connects to exactly three others—a parent and two children (Figure 1). This structure is simple and space-efficient because it limits the number of connections. However, it is also inflexible. If a node needs to connect to a different branch, it often can't, which leads to wasted nodes and inefficient layouts. Second a highly flexible, low scalability graph imagine a fully connected graph where every node can connect to every other node (Figure 2). This setup is highly flexible – it can represent any lambda Expression using the fewest possible nodes. But scalability is a major issue. Each time a node is added, a connection must be added to every other existing node, causing the total number of connections to grow exponentially.

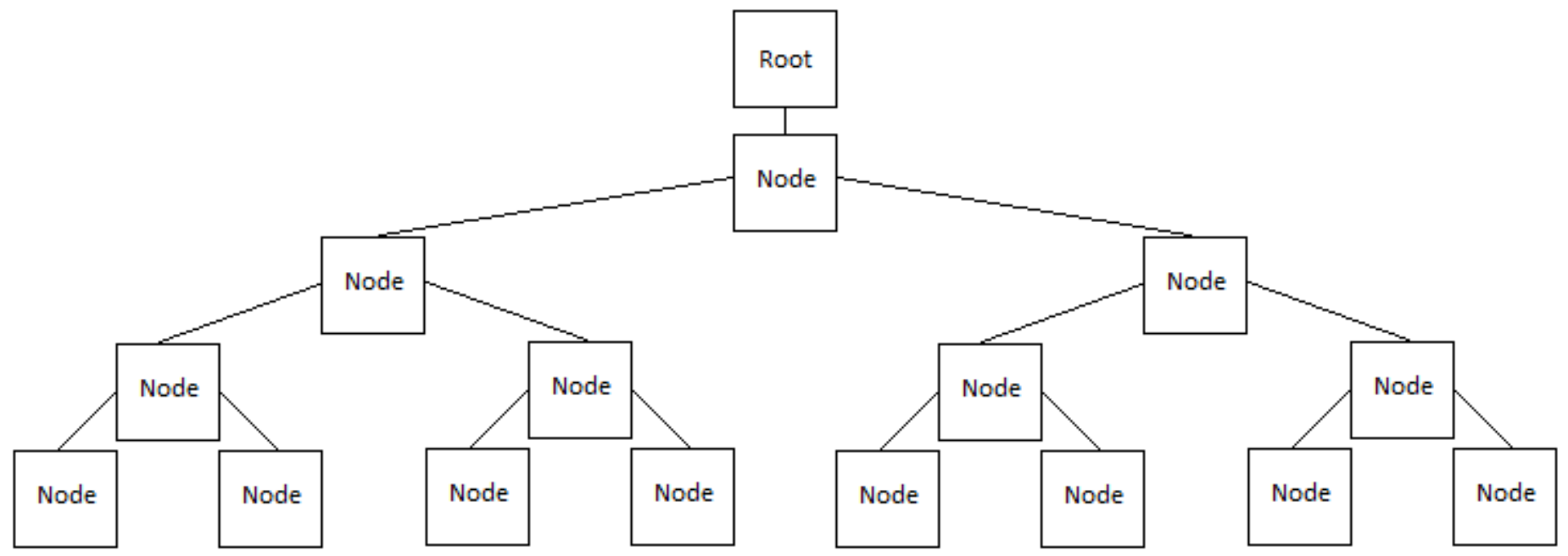

**Fig. 1**: Highly Scalable, Low Flexible Node Cluster

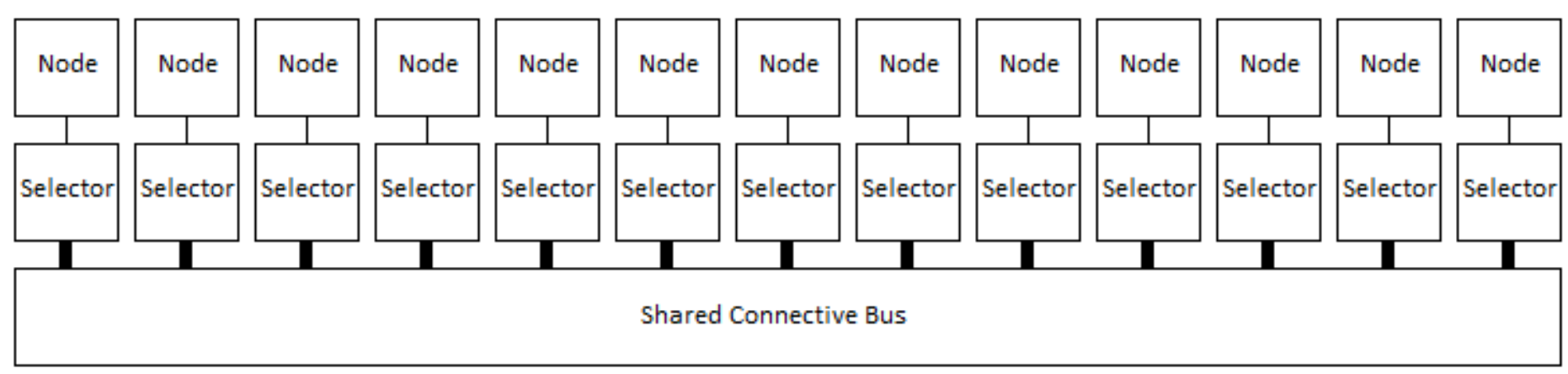

**Fig. 2**: Highly Flexible, Low Scalable Node Cluster

The approach used here aims to strike a balance between these extremes. Nodes are grouped into 'work clusters', each containing the same number of nodes. Within a work cluster, every node can connect directly to every other node. A few specific nodes in each cluster, the root node and child output nodes, can connect outside the cluster, enabling clusters to link together. These work clusters can then be grouped into larger units called super work clusters, which are also fully connected at the cluster level. This pattern continues hierarchically into hyper work clusters, and so on, scaling to accommodate the required number of nodes. This design offers a good compromise, it maintains enough flexibility for practical use without the unmanageable connection overhead of a fully connected graph. However, it does mean that some nodes in different clusters won't be able to communicate directly. An abstract visualization of this is shown in Figure 3.

## 2.3 Reusable Nodes

Graph transformations can sometimes leave groups of nodes disconnected from the main structure. To avoid wasting resources, these nodes need to be reset to a neutral state so they can be reused elsewhere. This is done through a process called nullification. A node can be instructed to nullify by its parent, and it propagates this instruction to its children (if any) before resetting itself. This works as a form of garbage collection, allowing clusters to clean up unused branches and reassign those nodes to new parts of the graph ensuring that all nodes remain useful and active where possible.

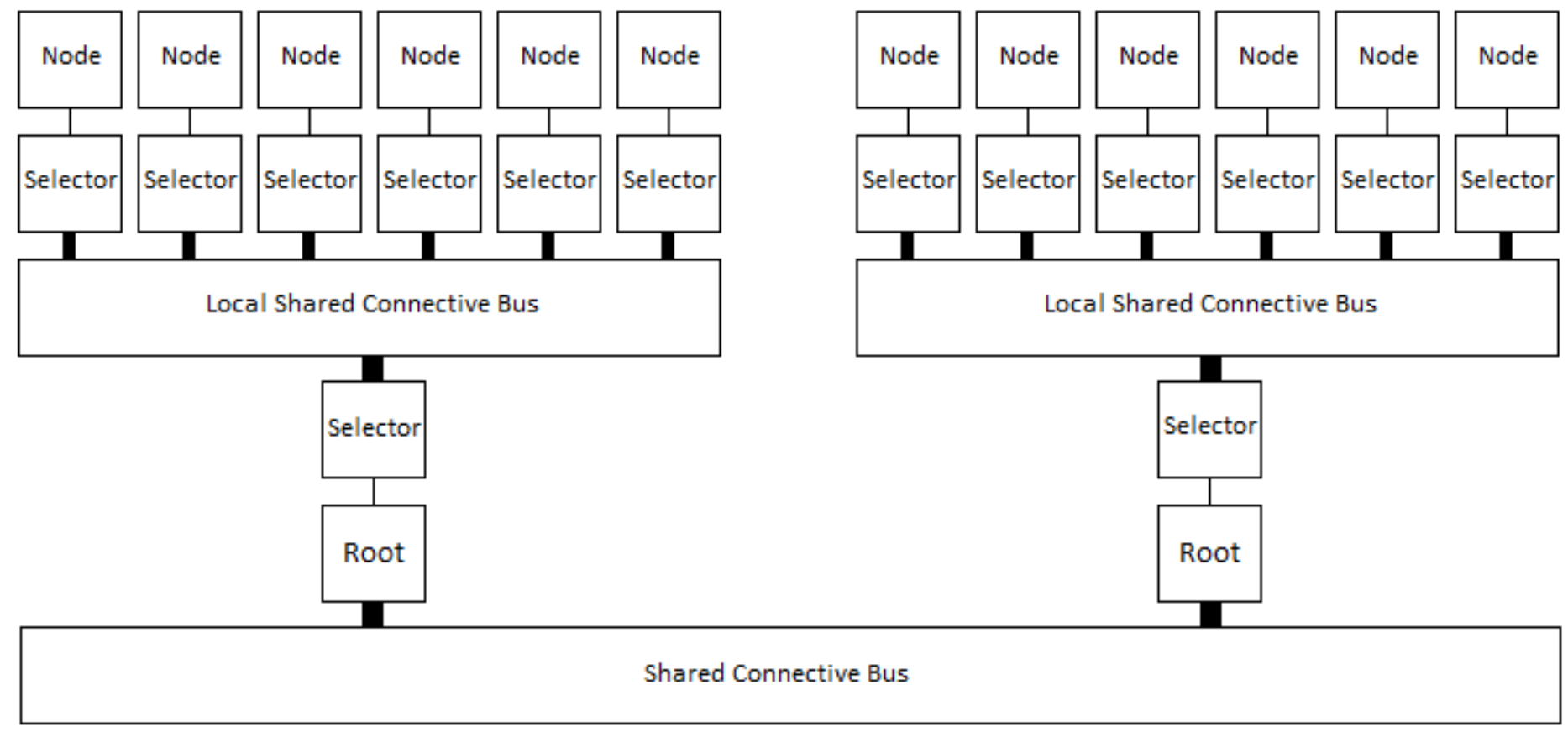

**Fig. 3**: Work Cluster

## 2.4 Implicit Alpha Conversion

An essential step before reduction in lambda calculus is alpha conversion. One or more alpha conversions may be required before some beta reductions can take place. Additional digital logic required to perform these transformations will be costly in both time and space. An alternate, more efficient solution to performing conventional alpha calculus exists due to the rigid and well-defined nature of digital logic.

Instead perform explicit syntactic alpha conversion at all, but have hardware act in equivalent but implicit way on the fly. The purpose of alpha conversion is to prevent multiple possible solutions existing for any Function due to shared Name values. For example, without alpha conversion, the expression $(\lambda^1 x.(\lambda^2 x.x)x)(\lambda^3 x.x)$ could be immediately reduced to two different expressions. In the first option, should the internal Function marked with a 2 be resolved, the expression reduces to $(\lambda^1 x.x)(\lambda^3 x.x)$ this is the correct solution. In the other option, where the external Function marked with a 1 is reduced initially, we are left with the expression $(\lambda^2(\lambda^3 x.x).(\lambda^3 x.x))(\lambda^3 x.x))$ which contains an invalid expression as the Function marked with a 2 now contains a non-Name figure next to the Lambda symbol. Alpha conversion would convert the expression to $^1(\lambda x.^2(\lambda y.y)x)^3(\lambda x.x)$, forcing the expression to reduce correctly. So we might say alpha conversion enforces a law that Functions should not reduce until all internal Functions have been reduced. However, sometimes Functions will not have inputs and are irreducible. For example, in the statement $(\lambda^1 x.(\lambda^2 x.x))(\lambda^3 x.x)$ the internal Function has no input and cannot reduce itself. Alpha conversion causes currently irreducible Functions to protect their contents by converting them to a new value. For example, $(\lambda^1 x.(\lambda^2 y.y))(\lambda^3 x.x)$. So alpha conversion enforces two laws: (1) that all internal reducible Functions must reduce themselves before the Function in which they are contained can reduce itself, and (2) Functions determined to be currently irreducible must protect their contents from transformation until the Function becomes reducible. So long as these laws are maintained, our model can prevent multiple solutions existing for any Function without performing explicit syntactic alpha conversion.

# 3 Data structure for expressions

Here we present a data structure for lambda expressions, to fit the hardware limitations above. It is a graph based structure which can be reduced in parallel, with nodes intended to be easily mappable to digital logic structures. We define the different expressions that a node can represent, the different values that are communicated, and the general functionality of each expression type.

## 3.1 Superclass Expression

Regardless of what expression nodes represent, they all share a number of attributes and methods, which can be captured by the object-oriented notion of a node superclass from which specific node types will inherit. nodes have the potential to connect to three other nodes: a parent node and two child nodes which we will refer to as child left and child right. Three attributes will contain these connected nodes: the parent pointer, the child left pointer and the child right pointer. There are 6 values that must be communicated between any two connected nodes:

First the graph as a whole requires a mechanism to determine if the graph can no longer be simplified. A boolean attribute called the *Resolve Flag* can be used to determine if all nodes on a branch below any given node currently have a raised Resolve Flag and thus no node on the branch will be performing any graph transformations. Using this, nodes can determine themselves to be resolved so long as their two children have a raised Resolve Flag, and when the root node has a raised Resolve Flag we can determine the graph to have been simplified.

Second as established in the previous section, some nodes that normally perform beta reduction may be determined to be *irreducible*. An expression will raise an *Irreducible Flag* to indicate that there is no valid Function input routed through it. Starting from the root node, a raised Irreducible Flag signal will be passed to child right. If a node ever receives a raised Irreducible Flag from its parent, it first determines itself to be irreducible, then propagates this signal to child right. This will cause all nodes on the rightmost branch of the graph to have raised Irreducible Flags. Additionally, some nodes may protect their contents from receiving inputs. For example, in the Function $(\lambda^1 x.(\lambda^2 y.y))z$, the internal Function expression marked with a 2 cannot access the z Name expression due to being encapsulated by the Function expression marked with a 1. To simulate this, expressions that encapsulate their contents should always pass a raised Irreducible Flag to child right.

Next the nodes will require an outbound and inbound connection for communicating instructional information. The type of send-receive behavior will depend entirely on which expression a node represents. Some expressions may produce instructions internally, and send them to connected nodes. Others may simply receive instructions from a connected node and pass them on to another connected node, effectively routing instructions between nodes. Others may act on received instructions, before sending their own instruction to connected nodes. Instruction data is comprised of an instruction and a node ID which can be used to specify a target node. Not all instructions require a node ID, and in these cases the node ID value will equal 0.

Finally two connections are needed for sending and receiving expression information. Expression information accompanies instructional information and while an instruction triggers a transformation, the expression information contains the values used during the transformation. Expression information includes the state of a Resolve Flag, expression type, and the contents of the child left and right pointers. Again, depending on expression type, nodes may route, transmit or replace expression information. This is used for sending and receiving the values needed during graph transformation.

### 3.2 Undefined Expression

The first node type is the Undefined expression, representing nodes disconnected from the tree structure. If any information is stored in the parent, child left and/or child right pointer, it will be replaced with a NULL value. It may seem redundant, but it is common during simplification that nodes will be disconnected from the graph structure. In a hardware implementation, nodes cannot be deleted and a NULL value must overwrite pointer information to avoid these disconnected nodes flooding buses with irrelevant information causing data collisions. These nodes where possible will be reinserted into the graph structure and may transform themselves into a different expression if appropriate instructions are received.

### 3.3 GoTo Expression

During graph transformations, for example when a Function expression is reduced, it must be replaced by a GoTo node. While not strictly necessary for a software implementation, this expression type is necessary for non-point-to-point implementations in digital logic. Its two purposes are to remove branches from a tree structure that need to be discarded, and to connect nodes together that may not have the ability to connect to each other. A GoTo node has a single child, all attributes and methods related to child left are not used. The need for GoTo nodes will be made clearer in section 5.

### 3.4 Name Expression

The Name expression is one of three expressions accessible to the programmer. Take the Function $(\lambda^1 x.(\lambda^2 y.y))z$ the values $x$,$y$ and $z$ are examples of Name expressions. Name expressions have a few properties that cause them to differ from the superclass node. First Name expressions are the only terminator nodes in our expression set, Meaning any branch will always end in Name nodes. Due to this, Name nodes can automatically and permanently raise their Resolve Flag. In conventional lambda calculus, Name expressions can store any alphabetic value from $[a - z]$. However, in digital logic, there's no such thing as an "alphabetic value." Instead, certain integer values can be converted into characters using an encoding system like ASCII. Since there's no practical reason to restrict values only to those that map to alphabetic characters, this approach allows any integer to represent a Name value. The Name expression is the only expression to store a value internally. All other expressions reference Name expressions when a value is needed. To accommodate this, an additional attribute will be required to store the name's value. Since Name nodes do not have children, Name information can stored in the attributes usually used as child pointers, as a union datatype (as in the C language).

Name values may replace themselves with another node type or expression value. During beta reductions, Names may have to change their internal value or change their expression type and expand the branch below itself. Details on how and when this is done will be provided in section 5.

### 3.5 Application Expression

The Application expression is the second expression accessible to the programmer. They are the main building blocks of the tree structure, perform no computation and, are primarily concerned with routing instruction and expression information between connected nodes. An Application can contain any expression, as either child, and acts as a branching point in the graph. For example, the Function $(\lambda^1 x.x)(\lambda^2 y.yy)$ contains two Applications. The first contains both functions, Function 1 being child left and and Function 2 being child right. The second Application is within the Function marked with a 2 and contains two Name expressions, both with values of $y$. Application routes information differently depending on the state of its Resolve Flag. An Application can determine itself to be resolved so long as both its children have their Resolve Flags raised. Figure 4 illustrates Applications in both these states.

Applications that have a raised Resolve Flag, meaning no nodes on the branch below them are reducible functions, must route instructions and expression data from potential Ancestor Function expressions above them, that at some point may attempt to perform a beta reduction, to Descendant Name nodes on the branch below. Instruction data received from both children is OR-ed together and passed to the parent node, as data may be received from either or both children. This allows Names to communicate to their Ancestor Functions when beta reductions have been completed. Relevant examples on when this happens will be shown in section 4. In the expression $(\lambda^1 x.x)(\lambda^2 y.yy)$, the Application containing two $y$-valued Names is an example of an Application in this state.

Applications with an unraised Resolve Flag must route expression data from a determined input node to a function, and route instruction data from the Function to the input. Despite any individual Application not being able to determine where on the tree an input and Function are, so long as all Applications route information in a standardized way, it is possible for Functions with a lowered Irreducible Flag to always receive their inputs no matter how many Applications are between them. To accomplish this, Applications in this state pass expression data received from their parent node to their child right node, and instruction data received from the parent to child left. Then expression data received from the child right node is passed to the child left node, and instruction data is passed to the Applications parent node. Finally, expression information received from the child left node is passed to the Application's parent node, and instruction information is passed to the child right node. For example, in $(\lambda^1 x.x)(\lambda^2 y.yy)$, the Function 1's input is the Function marked with a 2. Luckily they are both contained within an Application whose Resolve Flag is unraised. As an example the Application's child left is Function 1, and its child right is Function 2. So the Application passes expression data received from Function 2 to Function 1, and passes instruction data in the opposite direction, allowing the two nodes to communicate.

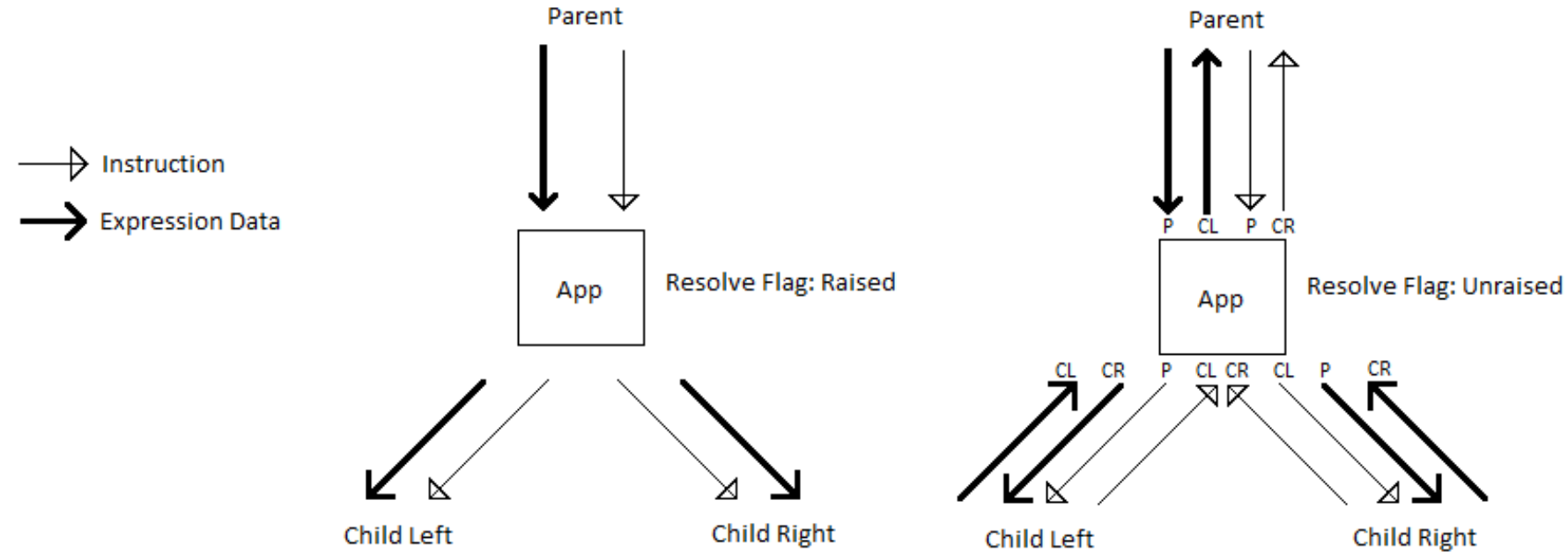

**Fig. 4**: Application Routing Examples

## 3.6 Function Expression

Functions are the final type of expression accessible to the programmer. Functions always have two children. The first child, child left, is always assumed to be, or reduce to, a Name. If not, the tree will simplify in a unpredictable manner likely causing errors. The second, child right, can be any expression. Functions coordinate beta reduction steps between the function's input and all applicable Name nodes. First, Functions must determine if they are reducible. If a function's parent node passes a raised Irreducible Flag that Function must raise its own Irreducible Flag. That Function is determined to be irreducible. Functions are assumed to be reducible unless the Irreducible Flag is raised. In both cases, regardless of whether a function's Irreducible Flag is raised, it must pass a raised flag to its child right. This prevents other Descendant Function nodes on the branch below the Function from retrieving input values not encapsulated by the function. A Function will also expect to receive expression information from its child left and parent nodes. The expression data is passed to child right, and instruction data is passed to both its parent and child right.

As outlined in section 2.4, Functions must comply with certain laws to implicitly perform alpha conversion. The first law applies to Functions that are considered reducible: all internal reducible Functions must reduce themselves before the Function in which they are contained can reduce itself. If a Function's children both have a raised Resolve Flag, that Function can determine that there are no Descendant Functions with lowered irreducible flags on the branch below itself. Additionally, a Function should check to see if its input is resolved. Any Function with a lowered Irreducible Flag can do this by reading the expression data sent by its input. This step is not entirely necessary as long as the input value is not mid-reduction when its connected Function begins reducing. Potentially a model can be made that allows Functions with a lowered Irreducible Flag to be used as inputs. However, this model avoids inserting reducible Functions as it potentially increases the number of graph transformations required to reach a solution, and more graph transformations requires more computation time. So after a Function with a lowered Irreducible Flag receives a raised Resolve Flag from its two children and its input, the Function can begin performing a beta reduction. To achieve this, a series of instructions are transmitted to the function's parent node and child right node, as explained in further detail in section 5.

The second law outlined in section 2.4 applies to Functions with a raised Irreducible Flag: Functions determined to be currently irreducible must protect their contents from transformation until the Function becomes reducible. If a function's parent passes a raised Irreducible Flag signal, and is therefore irreducible, upon receiving Resolve Flags from both of their children, the Function can raise its own Resolve Flag. This allows potential Ancestor Functions above it to begin reducing. However, some instructions, those related to performing beta reductions, received from parent nodes will never be passed to the function's children. More details on instructions, the expressions they effect and what they do are provided in section 4.2.

# 4 Node Intercommunication and Instructions

So far we have detailed how nodes exchange values through a series of shared buses. As established in section 3, nodes communicate to their parents and two children through an expression bus and an instruction bus. We will next show the various 'instructions' that nodes can send to one another to cause transformations. Instructions share similarities with instructions found in von Neumann architectures, but this should not be used to assume a global processing unit in control of all nodes within a cluster. A node may send an instruction to a connected node which may cause it to either update some of its internal values, return information and/or forward that instruction to another node.

(As with most hardware algorithms, it may be challenging for readers to follow details of algorithmic execution. To assist understanding, we provide a video demonstration covering the same algoriothms described here. See the Implementation section for details of this video, which can be watched together with the following descriptions.)

## 4.1 Node Defaults

Nodes store a series of internal values. All nodes within the graph are assigned a Unique Node ID which distinguishes that node from other nodes in the cluster. Three register exist for storing an expression type and two child pointers. Additionally, each node includes a stack to store other Unique Node ID's. Two variables front stack and back stack pointer will record the front of the stack, and the back of the stack respectively. Every node will have access to a temporary register that can contain an expression type called the expression buffer. Table 1 shows a list of all internal values and Table 2 defines buses nodes share between each other.

| Bus and Node Contents | Mnemonic |
|---|---|
| Internal Value | |
| Unique Node ID | UNI |
| Resolve Flag | RSF |
| Reduce Flag | RDF |
| Expression | EXR |
| Expression Buffer | EXB |
| Child Left Pointer | CLP |
| Child Right Pointer | CRP |
| Front Stack Pointer | FSP |
| Back Stack Pointer | BSP |
| Expression Bus Contents | |
| Resolve Flag | RSF |
| Expression | EXR |
| Child Left Pointer | CLP |
| Child Right Pointer | CRP |
| Information Bus Contents | |
| Instruction | INS |
| Unique Node ID | UNI |

**Table 1**: Value Definitions: Bus and Node Contents

| Node | Bus Data Type | |
|---|---|---|
| | Expression | Instruction |
| Parent | PEB | PIB |
| Child Left | CLE | CLI |
| Child Right | CRE | CRI |

**Table 2**: Value Definitions: Bus Types

Nodes act differently depending on which expression they are representing. Received instructions may cause the node to change its outputs or update its contents, however when no valid instructions are being received, they will default to a specific output pattern, which will be referred to as the 'expressions default'. Deriving from the definitions in section 3, Table 3 contains a complete collection of output patterns for expression defaults. Note that the '-' symbol can be interpreted as an empty output bus.

## 4.2 Instruction Set

Nodes communicate expression and instruction information to coordinate graph reduction. Instructions may be *Ancestor instructions* received from a parent, or *Descendant*

| Expression | | Output Value | | | | | |
|---|---|---|---|---|---|---|---|
| | | Parent | | Child Left | | Child Right | |
| | | PEB | PIB | CLE | CLI | CRE | CRI |
| Undefined | | - | - | - | - | - | - |
| GoTo | | [CRE] | [CRI] | - | - | [PEB] | [PIB] |
| Name | | {[RSF],[RDF],[CLP],[CRP]} | - | - | - | - | - |
| Application | RSF = 0 | - | - | [PEB] | [PIB] | [PEB] | [PIB] |
| | RSF = 1 | [CLE] | [CRI] | [CRE] | [PIB] | [PEB] | [CLI] |
| Function | | {[RSF],[RDF],[CLP],[CRP]} | [CRI] | - | - | [PEB] | [PIB] |

**Table 3**: Expression Defaults

*instructions* from a child. Ancestor instructions are used to coordinate graph transformations, while Descendant instructions, save for two instance, are reactionary markers, used to communicate when certain thresholds have been met. Instructions only affect specific expression types, causing their nodes to change stored values and output values from their defaults. Pseudocode examples are provided in algorithms 1-12 using keywords from Table 1.

### 4.2.1 Nullification

This solution may require new nodes to be added to the graph. To prevent previously discarded data from reappearing before a node can be reused, its contents must be nullified. This way when a new node is required the next available undefined node's Unique Node ID can be added as a child pointer value where the new branch is needed. When a node receives a nullification instruction from its parent it will pass it to all its children. This ensures every node on a branch will receive this instruction. Then the node sets its expression type to undefined and clears its child and stack pointers.

**Algorithm 1** Nullification, Affected Expression Types (GoTo, Name, Application and Function)

$[CRI] \leftarrow [PIB]$; $[CLI] \leftarrow [PIB]$;
$[EXR] \leftarrow [Undefined]$;
$[CLP] \leftarrow [0]$; $[CRP] \leftarrow [0]$;
$[SFS] \leftarrow [0]$;
$[SBS] \leftarrow [0]$

### 4.2.2 Return and Update

Nodes require instructions to update and retrieve the internal values of other nodes. This is used to setup/output the contents of a work cluster or to update/query branches during a graph transformation. To prevent multiple nodes being updated or returning simultaneously, these instructions also send the Unique Node ID of the node that needs to perform this instruction. When a node receives these instructions they are passed to its children then if the unique node ID received from the parent instruction bus matches its own unique node ID it returns or updates its internal values.

**Algorithm 2** UpdateExpression, Affected Expression Types (Undefined, GoTo, Name, Application and Function)

$[CRI] \leftarrow [PIB]; [CLI] \leftarrow [PIB];$
**if** $[PIB][UNI] == [UNI]$ **then**
    $[EXR] \leftarrow [PEB][EXR];$
    $[CRP] \leftarrow [PEB][CRP];$
    $[CLP] \leftarrow [PEB][CLP];$
    $[RSF] \leftarrow 0;$
**end if**

First the update instructions, `UpdateExpression`, shown in (Algorithm 2) causes recipient nodes matching the attached Unique Node ID to update their Resolve Flag, expression type and child pointer values to match the values received via the parent expression bus. Variations of this instruction exist for updating a specific pointer values seen in (Algorithm 4) and (Algorithm 3)

**Algorithm 3** UpdateChildLeft, Affected Expression Types (GoTo, Name, Application and Function)

$[CRI] \leftarrow [PIB]; [CLI] \leftarrow [PIB];$
**if** $[PIB][UNI] == [UNI]$ **then**
    $[CRP] \leftarrow [PEB][CRP];$
**end if**

**Algorithm 4** UpdateChildRight, Affected Expression Types (Name, Application and Function)

$[CRI] \leftarrow [PIB]; [CLI] \leftarrow [PIB; ]$
**if** $[PIB][UNI] == [UNI]$ **then**
    $[CLP] \leftarrow [PEB][CLP];$
**end if**

Next the instruction for returning an expression, `ReturnExpression` (Algorithm 5), cause nodes to return their expression information through the parent expression bus if their unique node ID matches or, if a match is found further down the branch. To accomplish this a Descendant marker instruction is sent to indicate which branch the desired node is on and thus which child expression bus to pass onward.

**Algorithm 5** Return Expression, Affected Expression Types (GoTo, Name, Application and Function)

$[CRI] \leftarrow [PIB]; [CLI] \leftarrow [PIB];$
**if** $[PIB][UNI] == [UNI]$ **then**
  $[PEB] \leftarrow \{[RSF], [EXR], [CLP], [CRP]\};$
  $[PIB] \leftarrow [(Mark)];$
**end if**

**if** $[CLI] == [(Mark)]$ **then**
  $[PEB] \leftarrow [CLE];$
  $[PIB] \leftarrow [CLI];$
**end if**
**if** $[CRI] == [(Mark)]$ **then**
  $[PEB] \leftarrow [CLE];$
  $[PIB] \leftarrow [CLI];$
**end if**

### 4.2.3 Branch and GoTo Chop

There are two Descendant instructions (that are not marker signals) used to add and remove GoTo nodes. During a graph transformation branches may need to be removed from the graph. This is done by converting Applications into GoTo nodes which chop off a child. The first instruction, `Branchchop` (Algorithm 6), causes Applications to send a nullification signal to the relevant child then update its expression type to become a GoTo node, chopping off the discarded branch in the process.

**Algorithm 6** Branch Chop, Affected Expression Types (Application)

$[EXR] \leftarrow [GoTo];$
**if** $[CLI] == [BranchChop]$ **then**
  $[CLI] \leftarrow [Nullify];$
  $[CLP] \leftarrow [CRP];$
**else**
  $[CRI] \leftarrow [Nullify];$
**end if**

Later GoTo nodes can remove themselves from the graph by sending a `GoToChop` (Algorithm 7) instruction to its parent, then nullifying itself. When a node receives a `GoToChop` instruction, it updates its child pointer to the equal the GoTo nodes child pointer. effectively jumping over the GoTo node in the process.

### 4.2.4 Graph Transformation Instructions

The following instructions relate to performing graph transformations. For a graph transformation to happen a Name or function, known as the Ancestor input, needs

**Algorithm 7** GoTo Chop, Affected Expression Types (Function, Application, GoTo)

**if** $[CLI] == [GoToChop]$ **then**
  $[CLP] \leftarrow [CLE][CRP]$;
**else**
  $[CRP] \leftarrow [CLE][CRP]$;
**end if**

to search its branch then another Name node, known as the Descendant input, needs to recreate the values being searched for. This is achieved through sending `ReturnExpression` and `UpdateExpression` instructions. However, while the expression produced by the Ancestor is directly copyable, new child pointers are required. To allow both nodes to keep track of a transformation, `TransformationPrepare` (Algorithm 8) is used to update the localized stack with the nodes that have been or will be queried. Anytime a node is queried, its front stack pointer (SFS) is increased by the number of children that the node's expression type should have; and the back stack pointer (SBS) is incremented by 1. Values X and Y are added to the stack, which are the Unique Node ID's of nodes that will be queried in the future. This way, a copy of the branch can be replicated any number of times, and both Ancestor and Descendant inputs can simultaneously yet independently know when the transformation has been completed by checking if the front stack pointer equals the back stack pointer.

**Algorithm 8** Transformation Prepare, Affected Expression Types (N/A)

$Input(X, Y, Z)$
**if** $[BSP]! = [FSP]$ **then**
  **if** $[PEB][EXP]! = [Name]or[GoTo]$ **then**
    $[RAM(FSP)] \leftarrow [X]$;
    $[RAM(FSP)] \leftarrow [Y]$;
    $[FSP] \leftarrow [FSP] + 2$;
  **end if**
  **if** $[PEB][EXP] == [GoTo]$ **then**
    $[RAM(FSP)] \leftarrow [X]$;
    $[FSP] \leftarrow [FSP] + 1$;
  **end if**
**else**
  $[PIB] \leftarrow [Z]$;
**end if**

### 4.2.5 Ancestor Transformation Instructions

There are two instructions that pertain to Ancestor inputs. The first `ImmediateResolution` (Algorithm 9) causes the Ancestor input to immediately send a `BranchChop` instruction to its parent, instantly chopping the Ancestor's branch off the graph.

The second, `AncestorTransformation` (Algorithm 10) which causes the Ancestor input to perform one of two actions depending on its expression type. If the input is a Name it returns its internal value to its parent through the parent expression bus (PEB). However, if the inputs is a Function it not only returns its internal value but methodically queries the full branch bellow itself. Then when the branch has been fully queried the Ancestor input sends a `GoToChop` instruction to its parent removing itself from the graph.

**Algorithm 9** Immediate Resolution, Affected Expression Types (Function, Name)

$[PIB] \leftarrow [BranchChop], [UNI]$

**Algorithm 10** Ancestor Transformation, Affected Expression Types (Function, Name)

$[CLI] \leftarrow [ReturnExpression][RAM(BSP)];$
$[CRI] \leftarrow [ReturnExpression][RAM(BSP)];$
**if** $[RAM(BSP)] == [UNI]$ **then**
  $TransformationPrepare([CLP], [CRP]], GotoChop);$
  $[PEB] \leftarrow \{[RSF][EXR][CLP][CRP]\};$
**else**
  **if** $[CLI] == [(Mark)]$ **then**
    $TransformationPrepare(\{[CLE][CLP]\}, \{[CLE][CRP]\}, GotoChop);$
    $[PEB] \leftarrow [CLE];$
  **else**
    $TransformationPrepare(\{[CRE][CLP]\}, \{[CRE][CRP]\}, GotoChop);$
    $[PEB] \leftarrow [CRE];$
  **end if**
**end if**
**if** $[SBS] == [SFS]$ **then**
  $[EXR] \leftarrow [EXB;]$
**end if**

### 4.2.6 Descendant Transformation Instructions

Two instructions are again required for Descendant inputs. First the Descendant inputs need to compare a value to check if a transformation is necessary, for example in the expression $(\lambda x.y)$, Name $y$ will not perform a transformation as it contains a different value to Name $x$. `CompareValue` (Algorithm 11) checks Name values, sends a marker signal to indicate an equivalence and marks the Name node as 'allowed for transformation'. The second required instruction is `DescendantTransformation` (Algorithm 12) which works in tandem with the `AncestorTransformation` instruction. The expression type of the first node received from the Ancestor is stored in the Descendants expression buffer. It then determines how many child nodes that expression type should have and retrieves the corresponding number of Unique Node IDs, adding them to its stack. As the Ancestor's branch is processed, the Descendant uses these IDs for `UpdateExpression` instructions. Each time an ID is used to update a node, it is removed from the stack. The Descendant determines the transformation has been complete when its stack is empty sending a marker signal to indicate the new branch has been created.

---

**Algorithm 11** Compare Value, Affected Expression Types (Name)

---

**if** $[EXP] == (Name)$ **then**
    **if** $([PEB][CLP] == [CLP] AND [PEB][CRP] == [CRP])$ **then**
        $[PIB] \leftarrow [(mark)];$ Allow for Transformation
    **end if**
**end if**
**if** $[EXP] == (Application)$ **then**
    **if** $[CLI][INS] == [(Mark)]$ **then**
        $[PIB] \leftarrow [CLI]$
    **else**
        $[PIB] \leftarrow [CRI]$
    **end if**
**end if**

---

**Algorithm 12** Descendant Transformation, Affected Expression Types (Name)

---

**if** $[RAM(Back)] == [UniqueNodeID]$ **then**
    $[EXB] \leftarrow \{[PEB][EXR]\};$
    $[CLP] \leftarrow [NewNode1];$
    $[CRP] \leftarrow [NewNode1];$
**else**
    $[CLI] \leftarrow \{[UpdateExpression][RAM]\};$
    $[CRI] \leftarrow \{[UpdateExpression][RAM]\};$
    $[CLE] \leftarrow \{\{[PIB][RSF]\}, \{[PIB][EXR]\}, [NewNode1], [NewNode2]\};$
    $[CRE] \leftarrow \{\{[PIB][RSF]\}, \{[PIB][EXR]\}, [NewNode1], [NewNode2]\};$
**end if**
**if** $[CLI] == [(Mark)]$ **then**
    $TransformationPrepare(\{[CLE][CLP]\}, \{[CLE][CRP]\}, [(Mark)]);$
**else**
    $TransformationPrepare(\{[CRE][CLP]\}, \{[CRE][CRP]\}, [(Mark)]);$
    $[SBS] \leftarrow [SBS] + 1;$
**end if**

---

# 5 Beta Reduction Example

This section will demonstrate a step-by-step beta reduction graph transformation. As an example, we will consider the expression $(x(\lambda^1 y.y))(\lambda^2 z.z)$, which reduces to $(x(\lambda^2 z.z))$.

## 5.1 Initialization

Figure 5a is an abstract visual representation of a graph structure representing the example expression $(x(\lambda^1 y.y))(\lambda^2 z.z)$. During transformation steps, buses containing information relevant to the transformation step will be highlighted to show how expression and instruction data is routed throughout the graph. This example begins mid-reduction. Function 2 is receiving an Irreducible Flag from its parent, and raised Resolve Flags from both children. This has caused it to raise its own Resolve Flag, and pass its internal values to its parent through the expression bus, which eventually arrives at Function 1. Function 1 is now receiving a raised Resolve Flag from both its children, and a raised Resolve Flag from its Ancestor input, Function 2. Beta Reduction can now begin. Function 1 instructs its second child to `CompareValue` expression data retrieved from its first child. If a marker signal is returned, Function 1 sends a `AncestorTransformation` instruction to the Ancestor input. Likewise, a `DescendantTransfromation` instruction is sent to all Descendant inputs, If Function 1 does not receive a `CompareValue` marker signal, it can send a `ImmediateResolution` to its Ancestor Input, and jump to the final step of beta reduction. Both inputs have readied themselves for transformation, by incrementing their front stack by 1, and appending their own unique node ID to their stacks.

## 5.2 Transformation Steps

Beta reduction can now begin the first step shown in Figure 5b. Function 1 continue to output Descendant/Ancestor transformation instructions, and will route its received parent expression bus to its child right expression bus, until it receives a marker indicating a resolution. The Ancestor input's stack currently contains one Unique Node ID: its own. This is pointed to by the back stack pointer. So the first node to be copied during transformation is Function 2, which will send the current state of its Resolve Flag, expression and child pointers to its parent. The Descendant input also updates the node pointed to by its back stack, which is currently itself. To prevent an expression change – which may cause synchronization errors – the Descendant input stores the expression type in its expression buffer, and will update its expression type later when the transformation is complete. Its child pointers are updated to point at two new undefined nodes. Next, both inputs add the Unique Node IDs of the children of the node just queried. For the Ancestor input, these are the Node IDs of the two 'Name: Z' nodes, and for the Descendant, they are the two new undefined nodes. Afterwards, the Ancestor input and the Descendant input both increment their front stack pointers by the number of children that the relevant expression type should have. In this example step, the expression type just queried was a Function which has two children. So both inputs update their front stack pointer (FSP) value from 1 to 3. Finally, the back stack pointer is increased by 1, and both inputs must check to see if the transformation has been completed. Currently, they both have a BSP of 1, and a FSP of 3, which are not equal, meaning the transformation is incomplete.

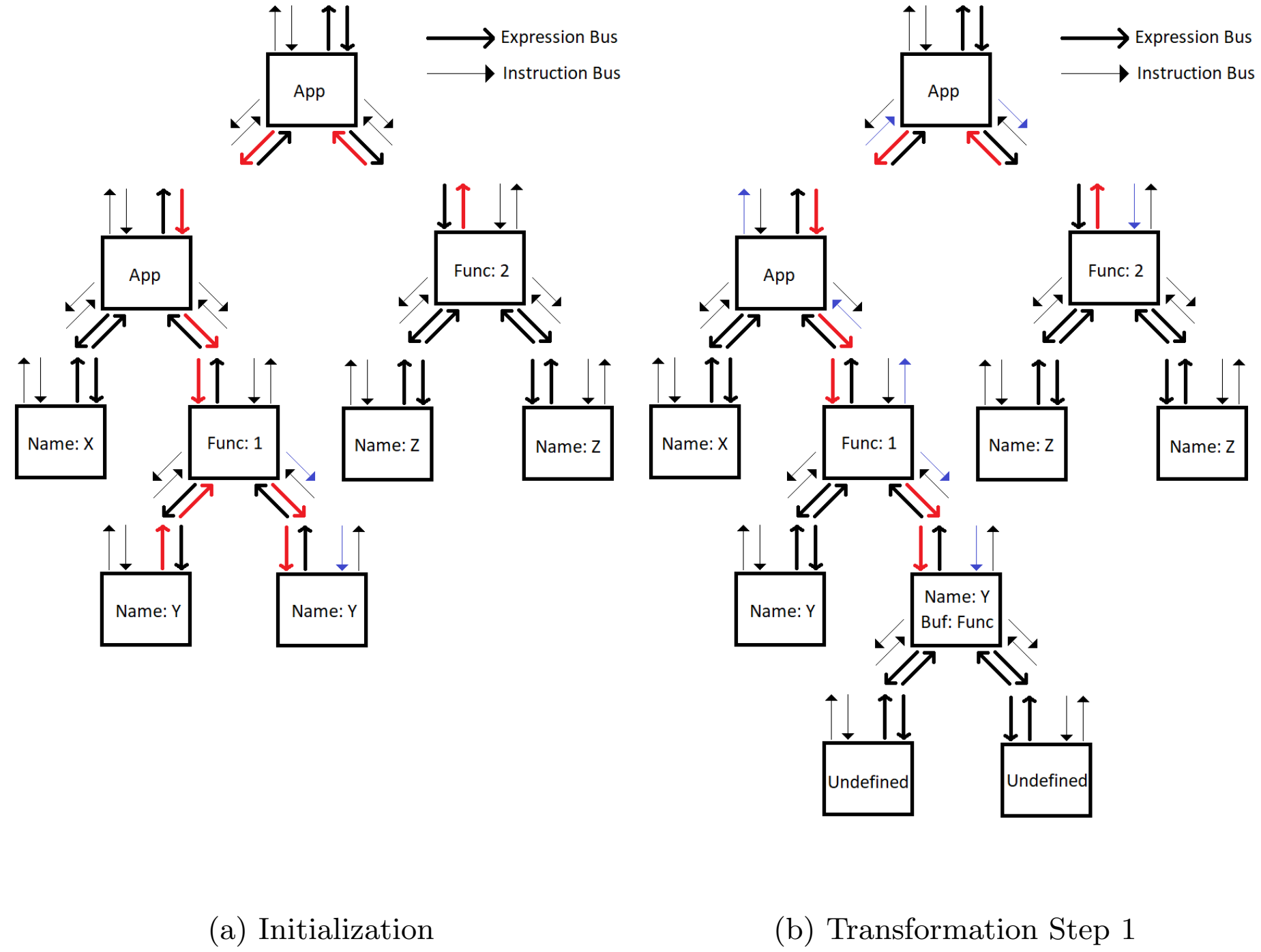

(a) Initialization

(b) Transformation Step 1

**Fig. 5**: Beta Reduction Initialization. Highlighted arrows represent buses containing information relevant to the current step

The second transformation (shown in Figure 6a) has both inputs query a node whose ID matches the back of their stack, as indicated by each input's back stack pointer. In this example, the Ancestor retrieves the contents of its leftmost child (Name Z) using the `ReturnExpression` instruction. The Descendant checks its stack, and also updates its leftmost child (currently undefined) using the `UpdateExpression` instruction transforming it into a Name node with matching contents. Again, both inputs go to update their front stack pointer. However, the expression most recently queried was a Name which has no child nodes. So neither the Descendant nor ascendant input increase their FSP or append any values to the stack. BSP is then incremented and with BSP and FSP still not equivalent a third transformation step begins.

The third and final transformation step, (shown in Figure 6b) of this example begins with both inputs querying the only remaining ID in their stacks pointed at by the BSP. In this case the rightmost child of both inputs. The final (Name Z) node is retrieved then copied into the last undefined node. Again, the front stack pointer does not increase as Names have no children. In both nodes, BSP increases to 3, which now matches their stored FSP value. Both inputs independently conclude that the transformation has been completed, and send markers/instructions to cause the resolution and nullification steps of a beta reduction.

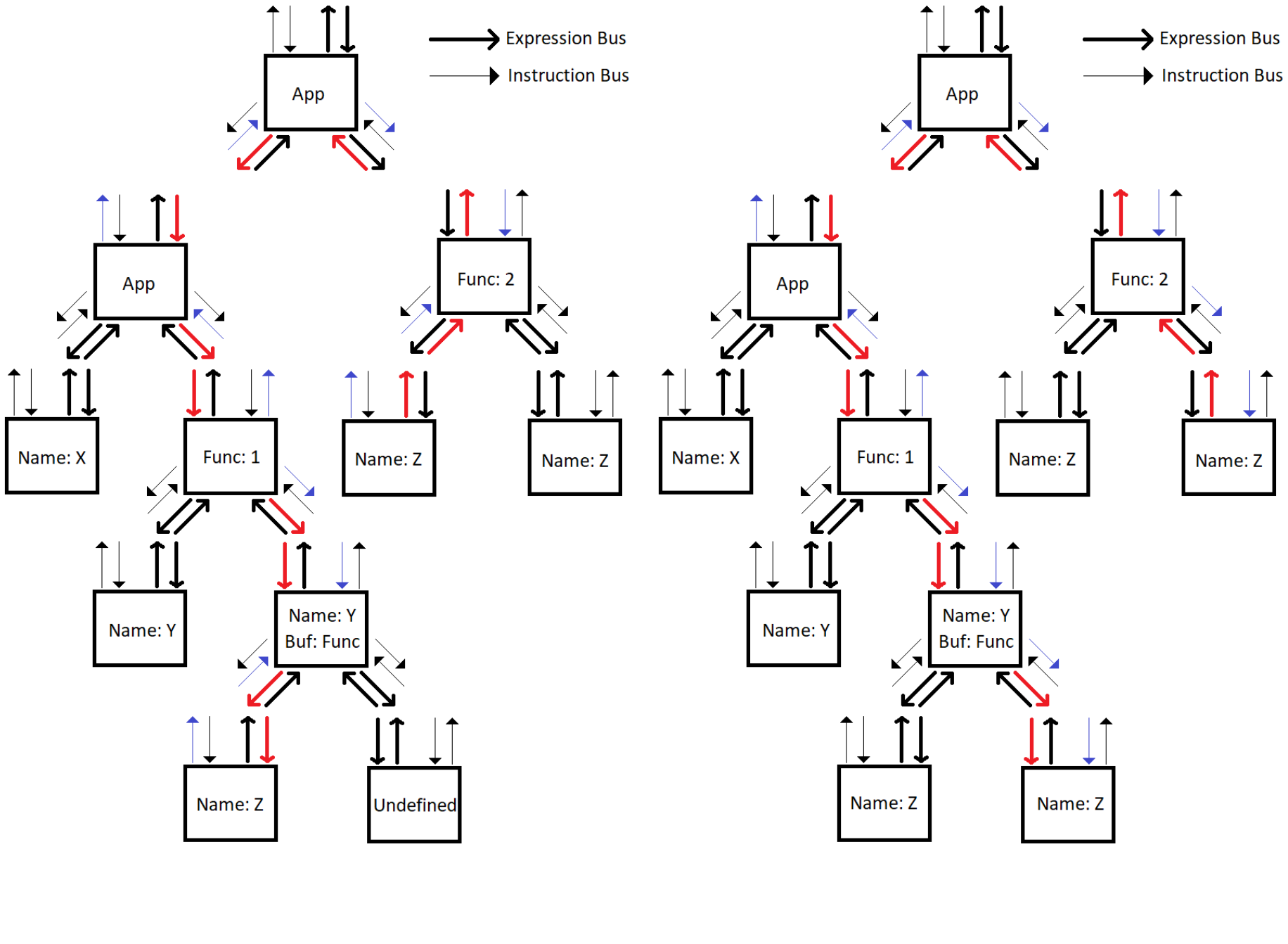

(a) Transformation Step 2 (b) Transformation Step 3

**Fig. 6**: Transformation Steps. Highlighted arrows represent buses containing information relevant to the current step

## 5.3 Nullification

Function 1, the Function that has just been reduced, has now successfully copied the Ancestor input's branch into the newly created Descendant input's branch and has received a marker signal from its Descendant input. The Function must now remove the compared Name value always found as the left-most child of any function. To accomplish this, the Function will nullify the undesired branch, then convert itself into a GoTo node. The nullification step involves Function 1 sending a Nullification instruction to its leftmost child.

The Ancestor input also needs to be removed from the graph. So sends a `BranchChop` signal and its own Unique Node ID to its parent node through the parent instruction bus. The parent node (an Application) needs to convert itself into a GoTo node by first nullifying the undesired branch. The `BranchChop` instruction compares the accompanying Unique Node ID with the Applications two child pointers and sends a Nullification instruction to the matching branch. This results in the branch containing the Ancestor input nullifying itself.

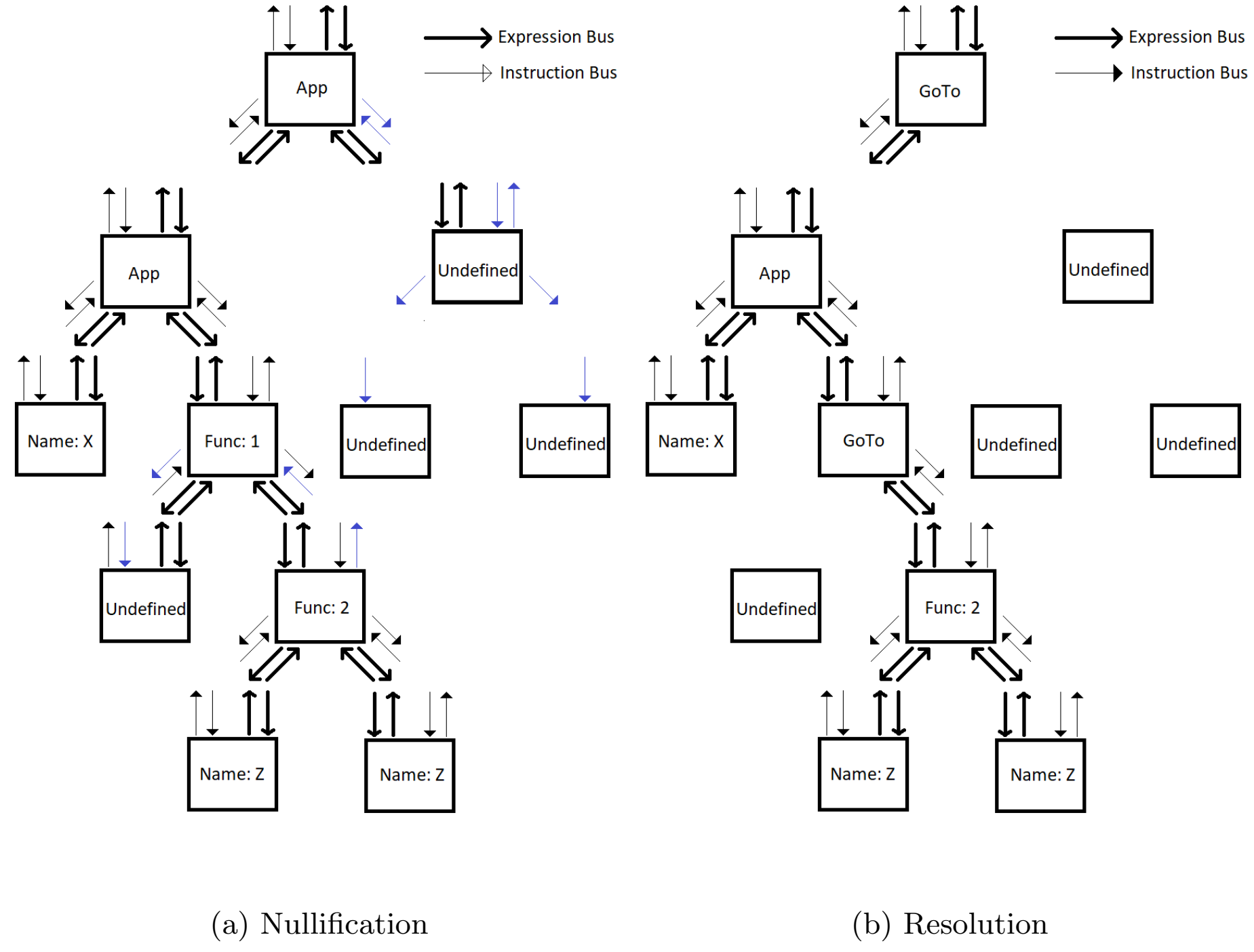

(a) Nullification

(b) Resolution

**Fig. 7**: Nullification and Resolution. Highlighted arrows represent buses containing information relevant to the current step

The Descendant input is currently still a Name node. However, it has two children, which is an invalid state for a Name node. To resolve this error the Descendant input must replace its expression type with the expression stored in its expression buffer register.

The nullification step of beta reduction is demonstrated in Figure 7a.

## 5.4 Resolution

Function 1 and the Application that was previously the parent of the Ancestor input update their expression type to become GoTo nodes. With this step, Function 1 has been fully reduced resulting in the graph shown in Figure 7b.

The undefined nodes have been properly disconnected and, if this graph were to perform another transformation, could be reinserted back into the graph. If a work cluster were to run out of undefined nodes during a transformation, the graph could be flooded with `GoToChop` instructions, and potentially reclaim more undefined nodes.

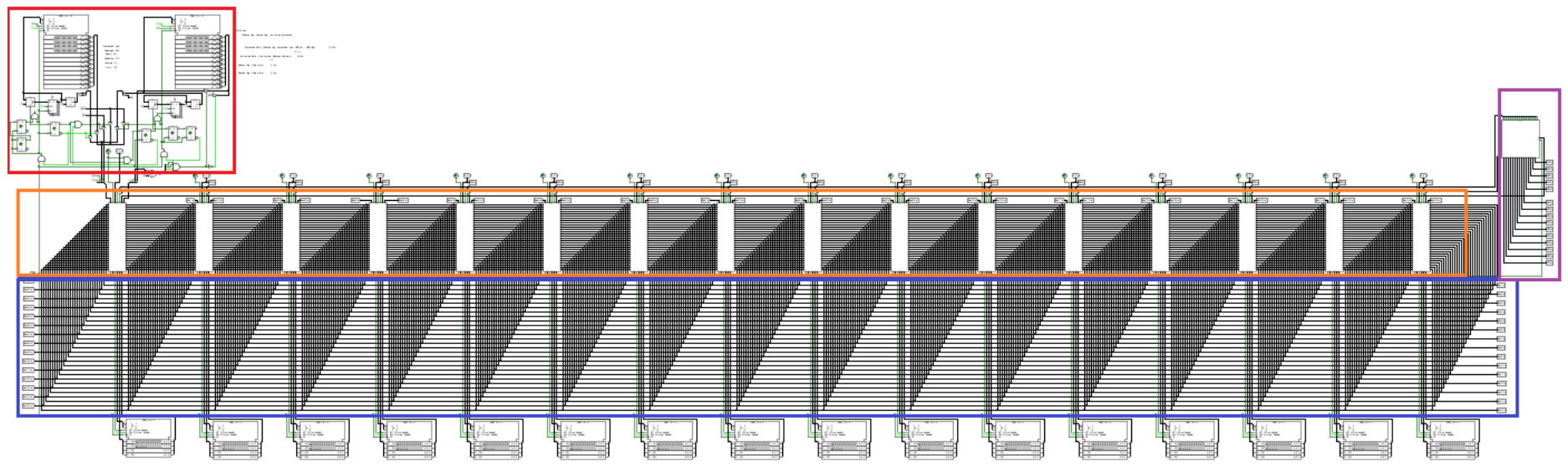

**Fig. 8**: Work Cluster Implementation

However, this process causes Functions to take longer to reduce, so GoTo nodes should only be reclaimed in the absence of undefined nodes.

To convert this graph back into a conventional lambda calculus expression assume that all GoTo nodes do not exist and their parents are directly connected with their children. Then trace the graph downwards from the root node to derive the expression. To demonstrate, in the example shown in Figure 7b, the graph begins with an Application pointing to two children a Name node storing an 'X' and a function. Continuing down the graph the Function has a further two children, both Name nodes storing a 'Z'. Therefore, this graph can be written as $(x(\lambda^2 z.z))$.

Following this reduction, Resolve Flags will propagate upwards and as there are no reducible Functions left the GoTo node at the top of the graph will eventually raise its Resolve Flag, passing this information to the work cluster's root node. Once this has happened, any external components reading the root nodes output can determine that there are no nodes within this work cluster that still require a transformation meaning the cluster has been reduced to its simplest possible form.

# 6 Implementation and Simulation

A work cluster has been implemented using Logisim Evolution and made available as open source[1]. This repository includes a copy of *logisim-evolution.jar*, the example execution file *ExampleExecution.txt*, and a circuit file *LambdaCalculus.circ* which exists in two versions. Version 1 is a lightweight version suitable for most expressions, however, this version is incapable of reducing some recursive or more complicated functions. Version 2 has an updated peripheral node tracker capable of managing multiple new node requests in a single clock cycle, enabling this version to handle any lambda expressions. Instructions for loading, reading, and reducing the sample execution are provided in the *README.txt* file in the repository. If a visual explanation is preferred, a video walkthrough of the work cluster is also provided[2].

Figure 8 shows the highest level of the work cluster. The majority of the figure is occupied by the 16 nodes that make up the cluster (highlighted with an orange box), all connected through a shared connective bus structure (shown in a blue box). Several peripheral components support the simulation. On the right side (in the purple box), a 'NewNodeTracker' component which keeps track of undefined nodes and provides new Node IDs during transformations. On the left side (in the red box), there are two RAM components along with some supporting logic. Since Logisim Evolution only allows example execution text files to be loaded into RAM components, the first RAM is used to automatically load example executions into the cluster. When the root node's Resolve Flag is raised, the simulation writes the updated cluster state into the second RAM. This setup allows users to compare the cluster's state before and after reduction. During execution, each node can be individually monitored. At the top of each node, an LED shows whether its Resolve Flag is active. There is also a 3-bit output that displays a value indicating the node's expression type. The provided encoder key (located near the input/output RAM components) can be used to decode these expression types. Each node's localized RAM component can be found at the bottom of the figure and during execution a user can monitor these to view each node's stack.

[1] https://github.com/LAMB-TARK/CPU-less-parallel-execution-of-Lambda-calculus-in-digital-logic
[2] https://www.youtube.com/watch?v=ch1fyDx1kn0

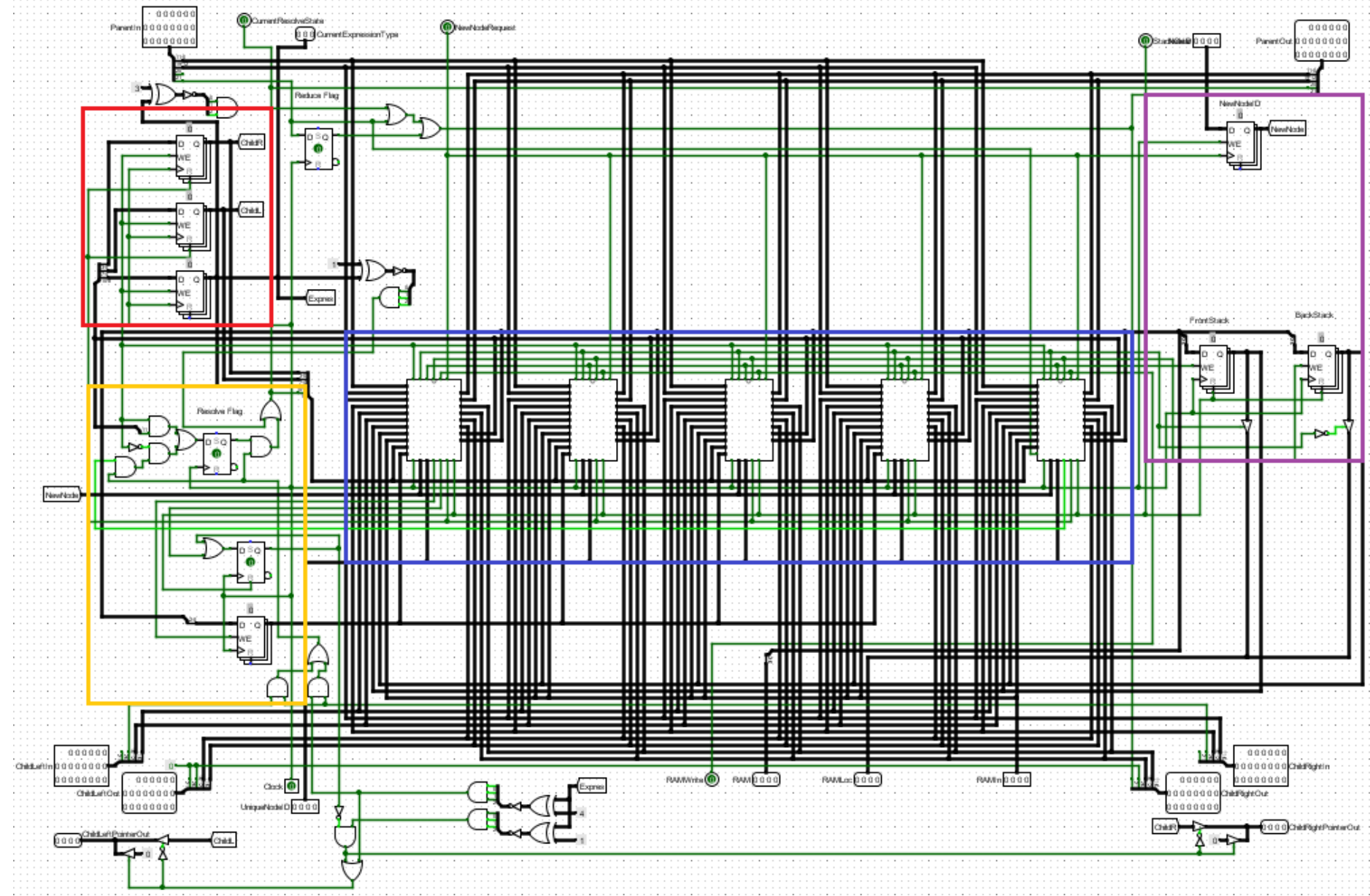

**Fig. 9**: Node Implementation

Figure 9 shows the internal structure of an individual node. A selector layer connects the node's inputs and outputs to its designated parent and child nodes, ensuring all pins are correctly routed. On the left side of the figure (in the red box) are the memory components that store key status values for the node. These include the expression register, as well as the left and right child pointers at the top. Lower down (in the orange box) is the Resolve Flag, expression buffer register and a buffer that toggles the selector layer depending on if a child pointer contains pointer information or a value. On the right side (in the purple box) are registers used for storing stack pointers and new node IDs. Note that the RAM components being referenced are external, found at the bottom of Figure 8). At the center of the node are five expression blocks, Highlighted in a blue box. These control the node's output values and may update the node's internal registers in response to specific instructions. The controlling expression block is determined by the current expression type.

| Index | Expression | Expected Reduced Expression | Nodes Used | Clock Pulses | Reduced Expression | Reduction Success |
|---|---|---|---|---|---|---|
| 1 | x | x | 1 | 0 | x | Success |
| 2 | xxxx | xxxx | 7 | 0 | xxxx | Success |
| 3 | $(\lambda x.x)y$ | y | 5 | 8 | y | Success |
| 4 | $(\lambda x.y)(\lambda z.z)$ | y | 7 | 23 | y | Success |
| 5 | x$(\lambda y.y)(\lambda z.z)$ | x$(\lambda z.z)$ | 9 | 23 | x$(\lambda z.z)$ | Success |
| 6 | $(\lambda x.x)(\lambda y.yy)$ | $(\lambda y.yy)$ | 13 | 61 | $(\lambda y.yy)$ | Success |
| 7 | $(\lambda x.x)(\lambda y.y)(\lambda z.z)$ | $(\lambda z.z)$ | 13 | 78 | $(\lambda z.z)$ | Success |
| 8 | $(\lambda x.x)(\lambda y.y)(\lambda z.z)(\lambda a.a)$ | $(\lambda a.a)$ | >16 | 78 | $(\lambda y.y)$ | Failure |
| 9 | $(\lambda x.xx)y$ | yy | 7 | 8 | yy | Success |
| 10 | $(\lambda x.xx)(\lambda y.y)$ | $(\lambda y.y)$ | 13 | 78 | $(\lambda y.y)$ | Success (V2) |
| 11 | $(\lambda x.xx)(\lambda y.yy)$ | – | – | Infinite | – | Partial (V2) |

**Table 4**: Test Expression and Results

# 7 Validation and Limitations

Table 4 presents a table of test lambda expressions. Each entry includes an index (used for reference), the expected reduced expression, and the actual reduced expression (used to evaluate reduction success). It also lists Nodes Used, the maximum number of nodes that were in use (i.e. nodes that did not represent a undefined expression) each clock pulse, and clock pulses, representing the number of pulses required for reduction excluding initialization and reading of the cluster. This section uses the data in Table 4 to explore how well the system handles parallelism and to highlight its current limitations. Note that expression 5 is the example expression described earlier in section 5.

Considering the limitations of two expressions that failed to reduce successfully: expression 8 and expression 11. First, expression 8. Although this expression produced a result within 78 clock pulses, the final output differed from the expected result. Investigation reveals that the expression attempted to use more than 16 nodes, exceeding the cluster's node limit. As a result, some Functions did not successfully copy their Ancestor inputs, causing data loss. This expression could easily be resolved by a work cluster with a larger cluster limit. Expression 7 (which uses Function chains) reduced correctly without exceeding the node limit, demonstrating that the solution can handle sequential reductions. If more nodes had been available, expression 8 would have reduced correctly. Next, expression 11, which is an infinitely reducible expression. Each time it is reduced, it produces a graph structurally identical to the original, meaning it has no stable reduced state. similar to a recursive Function lacking a base case. Technically, this solution was considered successful, as the root node's Resolve Flag was never raised, causing the cluster to continue waiting for a successful reduction indefinitely. However, each iteration produced an identical graph with the addition of two extra GoTo nodes. By the third iteration, the expression exceeded the cluster's node limit, causing subsequent reductions to produce a malformed but stable, irreducible graph. Since this resulted in a stable (albeit malformed) state, the solution ultimately failed to reduce infinitely. To support infinite reduction properly, a mechanism for GoTo garbage collection, invokable mid-reduction, needs to be devised. This test bench demonstrates that the system described in this document can reduce any non-infinite lambda calculus expression, given sufficient time and memory (i.e. available nodes).

Table 4 also demonstrates the implicit parallelisms this solution inherits from lambda calculus. For instance, even though expression 9 performs twice as many transformations as expression 3, both end up completing in the same number of clock cycles. This is because expression 9's transformations are processed in parallel. Initially, this trend seems to break for more complex inputs, expressions 5 and 10 have different reduction times. However, it's important to note that expression 10 is a recursive function. Its reduced form, $^1(\lambda y.y)^2(\lambda y.y)$, remains a reducible expression. When external interrupts are applied during simulation, both expressions perform their initial reduction in just under 23 clock pulses. The additional time taken by expression 10 is used for an extra transformation, which expression 5 does not require.

The test bench also demonstrates the solutions ability to perform garbage collection. Expression index 7 starts with a 11 node graph. Over the course of three sequential reductions, each adding two nodes, the graph should use a total of 17 nodes. Yet, the system manages all these transformations using only 13 nodes. This is achieved by nullifying nodes in removed branches, effectively recycling them for future use. As a result, this solution is capable of reducing long chains of expressions that it otherwise would not be able to reduce.

The solution inherits some limitations from lambda calculus. Specifically, reduction times can increase depending on the size of the Ancestor input. For example, expressions 3 and 6 reduce the same function, $^1(\lambda x.x)$ , but expression 6 has a much larger Ancestor input. As a result, expression 3 completes in just 8 clock pulses, 53 fewer than expression 6. While more efficient methods for expression graph reduction have been proposed, many are too resource-intensive or require excessive data exchange, making them impractical for digital logic implementations. Further research into more complex reduction methods is required.

# 8 Conclusion

The system is able to correctly reduce finite lambda calculus expressions in parallel, at the digital logic level. This suggests an approach to make use of increasing transistor density without higher clock speeds, by mapping functional program executions into parallel hardware.

However, as a practical programming language, the system inherits a number of flaws from lambda calculus, which was initially created for more theoretical purposes. The primary issue is the efficiency of Church's method for representing numbers, which is theoretically elegant but costly in practical computation time and node space, making this solution impractical for many algorithms. As is common in lambda-derived functional language design, increasing the number of expression types and methods would improve this. For instance, adding expression types to allow for list-like structures or branch prediction through future nodes [24], allowing expression types to perform arithmetic operations and logical comparisons and allowing Functions to apply multiple inputs to its contents either through multiple reduction iterations or allowing multiple simultaneous Ancestor inputs similar to Green style sugaring could greatly reduce computation time. However, more complex functionality requires more digital logic, increasing manufacturing costs, circuit size and energy requirements. Determining which expressions and methods should be added is a delicate balancing act and a future research question.

There are a number of limitations this solution will need to address in order to viably compete with modern hardware counterparts. Primarily, there is a setup/output bottleneck, so currently the only way to write to a node within a work cluster from an external source is through an `UpdateExpression` instructions passed to the cluster's root node, which is propagated through the graph. Due to the only external accessible point being the root node, only one `UpdateExpression` instruction can be sent per clock pulse. In our current simulation,to setup the largest possible graph using all 16 nodes within the cluster takes 16 clock pulse and this problem will worsen with larger node counts per work cluster. This issue is mirrored when reading the graph as each node's contents must be retrieved through a `ReturnExpression` instruction passed through the root node.

Furthermore while the initial statements that 'Digital logic gates are inherently parallel' and that '[functional languages] are easy to implicitly parallelize' are true. Our assumption that the combination of these two attributes would reduce the need for optimization compilation steps is not strictly correct. While our implementation implicitly reduces any given graph following the most efficient solution. A singular lambda calculus expression can be represented through multiple graphs which reduce differently. Our solution takes more time to compute unevenly weighted graphs where a single branch is significantly longer than any others. When compared to evenly weighted graphs, where each branch has a similar node depth, a digital logic solution to balance graphs pre-reduction would significantly increase computation time and circuit complexity. Therefore this solution will still likely require a programmer or compiler to optimize graph structures pre-reduction.

A more advanced, CPU-less hardware component taking inspiration from our proof of concept would be unlikely to replace von Neumann-descended hardware in the near future, due to the huge investments made into current CPU optimizations. But there may be some immediate applications where our comparatively simpler yet parallelized design could be advantageous. For example, in custom embedded systems, FPGA and low-cost ASICs [25], for digital signal processing on-board robots and other real-time, IoT fielded edge computing systems, where low latency, low cost, and low power consumption are needed.